\documentclass[9pt,conference,onecolumn]{IEEEtran}
\usepackage[utf8]{inputenc}
\usepackage[english]{babel}
\usepackage{graphicx}
\usepackage{amsmath}
\usepackage{tikz}
\usepackage{array}
\usepackage{booktabs}
\usepackage{float}
\usepackage{amssymb} 
\usepackage{mathrsfs} 
\usepackage{xargs}
\usepackage{siunitx} 
\usepackage{tikzscale}
\IEEEoverridecommandlockouts

\usepackage{mathtools} 
\usepackage{commath} 
\usepackage{hyperref}

\newcommand{\scriptedarrow}[1]{\ensuremath{\stackrel{\text{#1}}{\rightarrow}}}
\newcommand{\E}{\mathrm{E}}
\newcommand{\Var}{\mathrm{Var}}

\newcommand{\lefto}{\mathopen{}\left}
\newcommand{\righto}{\right}
\newcommand{\farg}[1]{\mathopen{}\left( #1 \right)}

\definecolor{darkblue}{RGB}{0, 5, 150}

\allowdisplaybreaks

\providecommand{\vectornorm}[1]{\left\lVert#1\right\rVert}

\providecommand{\vectornormbig}[1]{\big\lVert#1\big\rVert}

\title{Short Packet Structure for Ultra-Reliable Machine-type Communication: Tradeoff between Detection and Decoding}
\begin{document}

\author{\IEEEauthorblockN{Alexandru-Sabin Bana, Kasper Fløe Trillingsgaard, Petar Popovski, Elisabeth de Carvalho}
\IEEEauthorblockA{Department of Electronic Systems, Aalborg University, Denmark}\thanks{This work was supported in part by the European Research Council (ERC Consolidator Grant no. 648382 WILLOW) within the Horizon 2020 Program.}}
\maketitle
\setlength{\abovedisplayskip}{1pt}
\setlength{\belowdisplayskip}{0pt}
\begin{abstract}
Machine-type communication requires rethinking of the structure of short packets due to the coding limitations and the significant role of the control information. In ultra-reliable low-latency communication (URLLC), it is crucial to optimally use the limited degrees of freedom (DoFs) to send data and control information. We consider a URLLC model for short packet transmission with acknowledgement (ACK). We compare the detection/decoding performance of two short packet structures: (1) time-multiplexed detection sequence and data; and (2) structure in which both packet detection and data decoding use all DoFs. Specifically, as an instance of the second structure we use superimposed sequences for detection and data. We derive the probabilities of false alarm and misdetection for an  AWGN channel and numerically minimize the packet error probability (PER), showing that for delay-constrained data and ACK exchange, there is a tradeoff between the resources spent for detection and decoding. We show that the optimal PER for the superimposed structure is achieved for higher detection overhead. For this reason, the PER is also higher than in the preamble case. However, the superimposed structure is advantageous due to its flexibility to achieve optimal operation without the need to use multiple codebooks. 
\end{abstract}
\section{Introduction}

Machine-type communication (MTC) is central to the 5G systems, where it appears in two flavors: massive MTC (mMTC), focused on serving a large number of devices and ultra-reliable low-latency communications (URLLC), focused on serving a small number of devices with stringent latency and reliability constraints. The main vehicle of MTC is transmission of short packets, especially critical for achieving low latency \cite{pkt_structure_urllc}. 
Short packet MTC is significantly affected by the control information \cite{towards_massiveURLLC_short_pkts}, which can be reduced through 5G grant-free access \cite{URLLC_principles_magazine}, where uplink (UL) short packets are sent in specific resources, without prior device scheduling. 

Packet detection is the key auxiliary procedure that needs to be carried out with very high reliability when URLLC is considered. Detection and synchronization studies date back decades in the communications context \cite{optimum_frame_sync_massey}. 
Recent information-theoretic works have investigated the achievable tradeoff between the rate of reliable communication and the asynchronism exponent \cite{comm_strong_async_tcham}. It has been shown that up to an asynchronism level, reliable communication is possible by joint detection and decoding \cite{asynchr_polyanskiy} in the discrete memoryless channel (DMC). Furthermore, it has been shown that it is sub-optimal to optimize the detection and decoding separately~\cite{codeword_or_noise}, and that there is a tradeoff between the probabilities of false alarm, misdetection and decoding error.
Joint detection and synchronization for a DSSS system using differential encoding is treated in \cite{differential_preamble_detection}, where the probabilities of false alarm and misdetection are derived for the AWGN channel, and further extended to a multipath Rayleigh fading channel. The false alarm is considered to be impacting only if it occurs at a time $t<t_r$ before the actual packet arrival, where $t_r$ is the time required to recover after a false alarm event. 
A comparison of preamble and superimposed packet structures was done in \cite{superimposed_ofdm_wlan} for channel frequency offset (CFO) training in OFDM WLAN, where the robustness of the superimposed case was improved through a variable data rate for the information symbols. It was shown that for the same packet length as the preamble scenario, the superimposed structure obtains similar CFO estimation performance, while reducing the data decoding error probability. 
In \cite{low_latency_UR5g_finite_blocklength}, the authors show that there is a tradeoff between bandwidth, latency, reliability, and rate for short packets, using the information-theoretic results on finite-blocklength regime from \cite{finite_blocklength_polyanskiy}. The paper outlines the optimal way to exploit the available spatial and frequency diversity under a latency-reliability constraint. The interaction between error-control coding and channel estimation for a short packet scenario in an AWGN channel with unknown, constant gain over a block is investigated in \cite{liva_mismatched_csi}, where it is shown that for a single-antenna receiver, there is an optimal training length for which the required signal-to-noise ratio (SNR) is minimized.

These results motivate our interest in the optimal tradeoff between detection overhead and decoding of a short packet, transmitted in a point-to-point system with acknowledgement (ACK), under strict latency-reliability constraints as in URLLC. 
Two packet structures are analyzed. The first one is the time-multiplexed structure, in which a fraction of the Degrees of Freedom (DoFs) is used as a preamble for detection and the remaining for data transmission. The second structure uses all available DoFs for detection and data. We treat the specific instance in which the detection sequence and data are superimposed. We derive the probabilities of false alarm and misdetection, and show the tradeoff between detection and decoding by numerically determining the optimal overhead that minimizes the packet error probability (PER), making use of results from finite blocklength information theory \cite{finite_blocklength_polyanskiy}. The results show that there is a tradeoff in allocating resources between detection and decoding. The preamble case achieves the minimum PER for a smaller overhead than the superimposed structure, which implies that more resources are needed for detection in the superimposed case. Therefore, the overall packet error achieved is slightly worse for the superimposed structure. However, from a pragmatic perspective, the superimposed structure offers an enhanced adaptivity, as it can operate optimally by simply controlling the power allocation for detection, rather than changing the preamble length and coding rate.
\section{System model}

We consider the case of a point-to-point round-trip exchange consisting of the transmission of a short packet of $b$ bits and reception of a positive or negative ACK under stringent latency-reliability constraints.
The transmitter and the receiver are assumed to have established synchronization at the symbol level. The receiver does not know the precise packet arrival time  $\tau$, which must be estimated prior to attempting to decode the data. Furthermore, we assume that upon the generation of a packet, a strict deadline for reception of the acknowledgement at the device is enforced.

\begin{figure}[t]
    \centering
    \includegraphics[width=0.75\columnwidth]{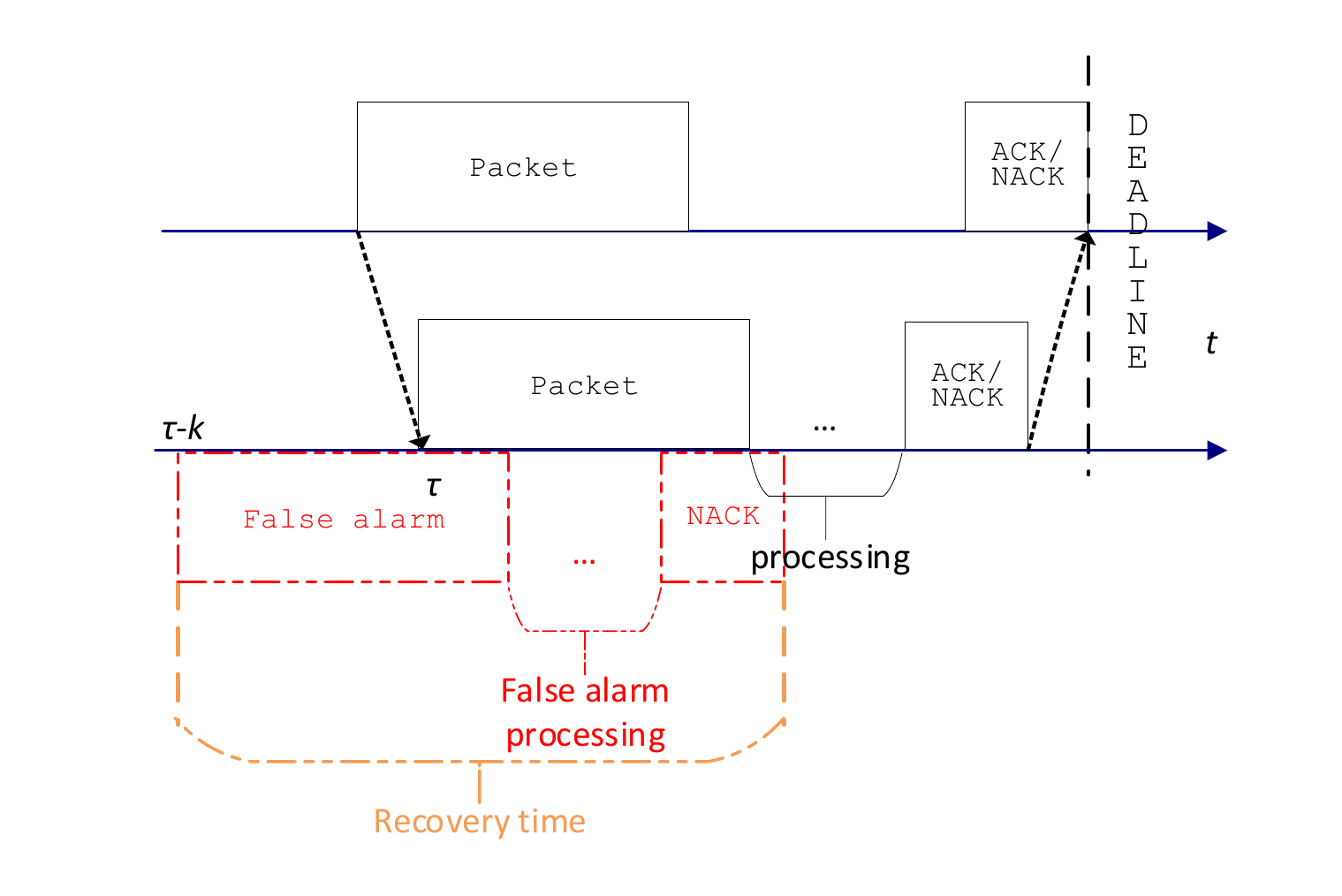}
    \caption{The system model considered in this paper. The two time axes represent the time lines at the transmitter and receiver, respectively. The model consists of a one-shot transmission with ACK/NACK feedback within a given latency constraint. The cost of having a false alarm is also shown, which means the loss of an incoming packet within the recovery time window.} 
    \label{fig:sys_model}
\end{figure}

A transmission is made in one shot, no retransmissions, and uses all the available resources: the sum of the DoFs (channel uses) spent for the packet transmission, (N)ACK, receiver processing, and the round-trip time amount to the maximum allowed latency, see Fig.~\ref{fig:sys_model}.
In such a system there are three main sources of uncertainty: the device activity, the noise, and the channel fading coefficients. 
Throughout this paper, the channel gain is assumed to be constant during a packet exchange and known by both the transmitter and the receiver, such that the channel input-output relation can be modeled as an AWGN channel, $Y_j = X_j + W_j$. 
Here, $\{W_j\}$ are i.i.d. circularly-symmetric complex Gaussian random variables with zero mean and unit variance and $X_j$ denotes the $j^{\text{th}}$ transmitted symbol, which is zero for $j \not\in \{\tau,\dots,\tau\mathopen{}+N\mathopen{}-1\}$. 
Assuming a fixed ACK structure and known  round-trip time, we focus on the transmitted packet structure, which contains the control information, the encoded $b$ information bits. We assume that the $b$ information bits include a cyclic redundancy check (CRC) which allows the receiver to detect errors. 
The round-trip PER can be expressed as
\begin{align}
    \mathcal{P}_e = 1-(1-\epsilon_{\text{d}})(1-\epsilon_{\text{D}})(1-\epsilon_{\text{ACK}})
    \label{eq:err_prob}
\end{align}
with $\epsilon_{\text{d}}$, $\epsilon_{\text{D}}$ and $\epsilon_{\text{ACK}}$ being the probabilities of error for detecting the packet, decoding the data, and correct ACK reception, respectively. As we aim to investigate the interplay between the detection and decoding, we assume $\epsilon_{\text{ACK}}=0$ and fixed number
of DoFs for ACK. This leaves $N$ 
channel uses for the detection sequence and the codeword that carries the data. Without loss of generality, we further abstract from the round-trip time for the remaining equations in the paper in order to improve the clarity of indexing. 

The receiver operates in a sequential mode: it can either run detection procedures and buffer samples of up to one packet length $N$ or transmit (N)ACK. Detection is performed using either time-multiplexed or superimposed Zadoff-Chu detection sequences \cite{chu_polyphase_codes}, \cite[Chapter 9]{LTE_for_4G_farooq_khan}, which have unit power per symbol and odd length. The root used to generate the sequence is chosen such that the partial-period correlation is minimal for each sequence length.
The coded information symbols, the noise, and the detection sequence symbols are further denoted by $D_j$, $W_j$, and $p_j$, respectively, where $p_j$ is zero for $j \not\in \{0,\dots,N_\text{p}-1\}$ for the preamble case, and for $j \not\in \{0,\dots,N-1\}$ for superimposed. We sometimes let the vectors $\mathbf{D}$ and $\mathbf{p}$ denote the coded information symbols and preamble sequence, respectively, with superscripts denoting their length.
The information symbols are assumed to be coded using spherical Gaussian codebooks (shell codes) for which the codewords are uniformly distributed on the shell of an $N_\text{c}$-dimensional sphere of radius $\sqrt{N_\text{c}}$, where $N_\text{c}$ is the codeword length which is $N-N_\text{p}$ for the preamble case and $N$ for the superimposed case. Such a codebook is capacity-achieving and achieves the optimal channel dispersion as well~\cite{finite_blocklength_polyanskiy}. We model the codewords as random vectors defined by
\begin{align}
    \mathbf{D}^{(N_\text{c})} = \sqrt{N_\text{c}} \dfrac{\mathbf{\tilde{D}}^{(N_\text{c})}}{\vectornormbig{\mathbf{\tilde{D}}^{(N_\text{c})}}}
    \label{eq:codeword_def}
\end{align}
where $\mathbf{\tilde{D}}^{(N_\text{c})} \sim \mathcal{N}(0,\mathbf{I}_{N_\text{c}})$ and $\vectornorm{\cdot}$ is the $\ell_2$-norm. The transmitted signals for the preamble and superimposed case, respectively, can be expressed as
\begin{align}
    X_{j+\tau}^{\text{(p)}} = \begin{cases} \sqrt{P} p_{j} , & j \in \{0,\dots,N_\text{p}-1\} \\
    \sqrt{P} D_{j-N_\text{p}}, & j \in \{N_\text{p},\dots,\mathopen{}N\mathopen{}-1\} \\
    0, & \mbox{otherwise} \end{cases}
    \label{eq:pream_y}
\end{align}
and
\begin{IEEEeqnarray}{ll}
    X_{j\mathopen{}+\tau}^{\text{(SI)}} \mathopen{}= \begin{cases}
    \sqrt{P} \farg{\sqrt{\alpha} p_{j} \mathopen{}+ \sqrt{1\mathopen{}-\alpha} D_{j}}, & j \mathopen{}\in \mathopen{}\{0,\dots,N\mathopen{}-1\} \\
    0, & \text{otherwise}. \end{cases} \IEEEeqnarraynumspace
    \label{eq:training_y}
\end{IEEEeqnarray}
Here,  $\alpha$ is the ratio of the powers spent for detection versus codeword for the superimposed case.

As we are interested in the regime where $N$ is small, we make use of the asymptotic approximations for the maximum coding rate for the AWGN channel, provided in \cite[Theorem 54]{finite_blocklength_polyanskiy}, to analyze the decoding error probability. It has been noted in \cite[footnote 1]{towards_massiveURLLC_short_pkts} that transmitting a length-$N_\text{c}$ codeword in a complex AWGN channel is equivalent to transmitting a codeword of length $2N_\text{c}$ in a real-valued AWGN channel, with the same SNR.
Therefore, the decoding error probability for a codeword of length $N_\text{c}$, carrying $b$ bits, given an SNR $P$ is well-approximated by 
\begin{align}
    \epsilon_\text{D}(N_\text{c},P) = \text{Q} \farg{\dfrac{2N_\text{c} C(P) - b + \frac{1}{2} \log_2 2N_\text{c} }{\sqrt{2N_\text{c} V(P)} } }
    \label{eq:error_finite_blocklength}
\end{align}
where ${C(P)= \dfrac{1}{2} \log_2(1+P )}$ and ${V\mathopen{}(P)=\dfrac{P(P+2)}{2(P+1)^2}} \log_2^2 e$ denote the channel capacity and dispersion, respectively.

The detection is performed by hypothesis testing, using the log-likelihood ratio. This is optimal in an AWGN channel and the metric of comparison is known as  \textit{deterministic correlation} \cite[Chapter 14]{MIT_intro}. 
Considering that a packet begins at time index $\tau$, the correlation at a time index $\tau-k$ can be expressed as 
\begin{align}
    \mathcal{R}_{Y,\tau-k} = \Re\mathopen{}\left[\sum\limits_{j=0}^{N_\text{t}-1} p_{j}^*  Y_{j+\tau-k} \right] > \Delta \label{eq:corr_metric}
\end{align}
where $\Re[x]$ denotes the real part of $x$, $p_j^*$ is the complex conjugate of the $j^\text{th}$ preamble symbol, $N_\text{t} \in \{N_\text{p},N\}$ denotes the length of the preamble and superimposed detection sequence, respectively, and $\Delta$ is the detection threshold.
Three error events can occur: (1) 
the false alarms ($\mathcal{E}_{\text{FA}}$) can occur at any time index ${\tau-k}$ if $\mathcal{R}_{Y,\tau-k}>\Delta$ for any offset $k \in \{1,\dots,t_r-1\}$, where $t_r$ denotes the \textit{recovery time} \cite{differential_preamble_detection} elapsed decoding, verifying the CRC bits, and transmitting NACK, during which the receiver misses any incoming packet (Fig.~\ref{fig:sys_model}). 
The receiver was assumed active for at least $t_r$ time slots prior to a packet arrival, such that $\tau-k>0$; 
(2) the misdetection event ($\mathcal{E}_{\text{MD}}$) occurs if $\mathcal{R}_{Y,\tau}\leq\Delta$;
(3) erroneous decoding event ($\mathcal{E}_{\text{D}}$), signaled by the CRC, due to a bad noise realization and the receiver sends NACK.
For the superimposed case, once the detection is in place, the detection sequence is subtracted from the received signal, such that it does not interfere with the decoding. 

Our model for the AWGN channel and URLLC is different from the other models in the literature \cite{comm_strong_async_tcham}, \cite{asynchr_polyanskiy}, \cite{codeword_or_noise}, which are valid for the discrete memoryless channels (DMCs) and treat an exponential level of asynchronism. This paper highlights that for URLLC in the discrete AWGN channel, it is important to model the receiver recovery time after a possible false alarm. This is especially true because for URLLC with ACK, there is an inherent cost to having false alarms, resulting in loss of the packet. 

\section{Analysis}

Using the three previously defined error events, we can formulate an upper bound on the PER as
\begin{IEEEeqnarray}{l}
    \mathcal{P}_{e}\leq \Pr[\mathcal{E}_{\text{FA}}] + \Pr[\mathcal{E}_{\text{MD}}] + \Pr[\mathcal{E}_{\text{D}}]. \IEEEeqnarraynumspace
\label{eq:probab_err_upper}
\end{IEEEeqnarray}
An approximation of the last term in \eqref{eq:probab_err_upper} is given by $\epsilon_\text{D}(N-N_\text{p},P)$  for the preamble case and by $\epsilon_\text{D}(N, (1-\alpha)P)$ for the superimposed case. The objective is to estimate the remaining two terms, namely the false alarm and misdetection probabilities and identify the parameters that influence the upper bound for both packets structures. 
\subsection{Time-multiplexed preamble}

We first analyze the false alarm probability $\Pr[\mathcal{E}_{\text{FA}}]$. A false alarm occurs if $\mathcal{R}_{Y,\tau-k}>\Delta$ for some $k \in \mathcal{S}_\text{FA}=\{1,\dots,t_r\mathopen{}-1\}$. 
The correlation output when a false alarm may occur is then
\begin{align}
    \mathcal{R}_{Y,\tau-k} = \sqrt{P} \mathcal{R}_{\text{p}}^\text{pre}(k) + \Re\mathopen{}\left[\sum\limits_{j=0}^{N_\text{p}-1} p^*_{j} W_{j+\tau-k}\right], \label{eq:R_te}
\end{align}
where $\mathcal{R}_{\text{p}}^\text{pre}(k)$ is the partial-period correlation of the preamble sequence, which is deterministic and nonzero for $k \in \{1,\dots,N_\text{p}-1\}$
\begin{align}
    \mathcal{R}_{\text{p}}^\text{pre}(k) = \Re\mathopen{}\left[\sum\limits_{j=k}^{N_\text{p}-1} p^*_{j}p_{j-k}\right]. \label{eq:rp}
\end{align}
The second term in \eqref{eq:R_te} is distributed as a Gaussian random variable. Thus, the probability of having a false alarm occurring at $\tau-k$, for any $k \in \mathcal{S}_\text{FA}$, is given by 
\begin{IEEEeqnarray}{rCl}
    \Pr\mathopen{}\left[\mathcal{R}_{Y,\tau-k} \mathopen{}> \Delta \right] &=& \Pr\mathopen{}\Bigg[ \sqrt{P} \mathcal{R}_{\text{p}}^\text{pre}(k) + \Re\bigg[\sum\limits_{j=0}^{N_\text{p}-1} p^*_{j} W_{j\mathopen{}+\tau\mathopen{}-k}\bigg] \mathopen{}> \Delta \Bigg] \nonumber \\
    &=& \text{Q} \mathopen{}\bigg( \dfrac{\Delta - \mu_{\mathcal{R}_{Y,\text{FA}}}(k)}{\sigma_{\mathcal{R}_{Y,\text{FA}}}} \bigg).
    \IEEEeqnarraynumspace
\end{IEEEeqnarray}
Here, the mean and variance of the correlation metric are
\begin{IEEEeqnarray}{rll}
    \mu_{\mathcal{R}_{Y,\text{FA}}}(k) &= \sqrt{P} \mathcal{R}_{\text{p}}^\text{pre}(k) \label{eq:mu_pre}; \IEEEeqnarraynumspace \\ 
    \sigma_{\mathcal{R}_{Y,\text{FA}}}^2 & \mathopen{}= \sum\limits_{j=0}^{N_\text{p}-1} \Var\mathopen{}\left[ \Re\mathopen{}\left[p^*_j W_{j\mathopen{}+\tau\mathopen{}-k} \right]\right] \mathopen{}=  \dfrac{N_\text{p}}{2} \label{eq:sigma_pre}
\end{IEEEeqnarray}
where \eqref{eq:sigma_pre} follows because the power of the Zadoff-Chu sequence $p_j$ is $1$ for $j\in\{0,\dots,N_{\text{p}}-1\}$.
Furthermore, the total probability of false alarms is the probability of the union of such events: 
\begin{align}
    \Pr\left[\mathcal{E}_\text{FA}\right] = \Pr\Bigg[ \underset{k\in \mathcal{S}_\text{FA}}{\bigcup} \mathopen{}\left\{ \mathcal{R}_{Y,\tau-k} > \Delta \right\}\Bigg]
    \leq \sum\limits_{k=1}^{t_r-1} \text{Q}\bigg(\dfrac{\Delta - \mu_{\mathcal{R}_{Y,\text{FA}}}(k)}{\sigma_{\mathcal{R}_{Y,\text{FA}}}} \bigg).
    \label{eq:P_fa2}
\end{align}

The misdetection error event $(\mathcal{E}_\text{MD})$ occurs if upon a packet arrival, $\mathcal{R}_{Y,\tau} \leq \Delta$. The probability of a misdetection is then
\begin{align}
    \Pr[\mathcal{E}_\text{MD}] &= \text{Q} \bigg (\dfrac{ \mu_{\mathcal{R}_{Y,\tau}} - \Delta}{\sigma_{\mathcal{R}_{Y,\tau}}} \bigg) \label{eq:misdetect}
\end{align}
where

\begin{align}
     \mu_{\mathcal{R}_{Y,\tau}} &= \E\bigg[ \sum\limits_{j=0}^{N_\text{p}-1} \Re\left[\sqrt{P}p^*_j p_j \right]\bigg] = \sqrt{P} N_\text{p}; \\
    \sigma_{\mathcal{R}_{Y,\tau}}^2 &= \sum\limits_{j=0}^{N_\text{p}-1}\Var\mathopen{}\bigg[\Re\left(p^*_j W_j\right)\bigg] = \dfrac{N_\text{p}}{2}. \label{eq:sigma_pre2}
\end{align}

\subsection{Superimposed sequence}

Similarly to the preamble case, we first derive the false alarm probability. For the superimposed case, the correlator output is given by
\begin{IEEEeqnarray}{l}
\mathcal{R}_{Y,\tau-k} = \sqrt{\alpha P} \mathcal{R}_\text{p}^\text{SI}(k) \mathopen{} +\sqrt{(1\mathopen{}-\alpha)P}\mathcal{R}_\text{D}(k) + \underbrace{\Re\bigg[\sum_{j=0}^{N\mathopen{}-1} p_{j}^* W_{j\mathopen{}+\tau-k} \bigg]}_{\mathcal{R}_W(k)}\mathopen{}, \IEEEeqnarraynumspace
\end{IEEEeqnarray}
where $\mathcal{R}_\text{D}(k) = \Re\bigg[\sum\limits_{j=k}^{N-1} p_j^* D_{j-k}\bigg]$ is the partial correlation between the detection sequence and the codewords. Due to the additional random variable $\mathcal{R}_\text{D}(k)$, we distinguish between two types of false alarms, occurring: (1) purely due to noise $(\mathcal{E}_{\text{FA1}})$ and (2) also due to partial-correlation with the preamble and codeword symbols $(\mathcal{E}_{\text{FA2}})$. The distributions of the correlation for the two types will be different. 
The partial period correlation of the detection sequence, $\mathcal{R}_\text{p}^\text{SI}(k)$, is defined similarly to \eqref{eq:rp} and remains deterministic. 

For the purely noise-inflicted false alarms, occurring when ${k \in \{N,\dots,t_r-1\}}$, both $\mathcal{R}_\text{p}^\text{SI}(k)$ and $\mathcal{R}_\text{D}(k)$ are equal to zero. Therefore, the noise-inflicted false alarm probability is
\begin{align}
    \Pr\left[\mathcal{E}_\text{FA1}\right]&\leq (t_r\mathopen{}-N) \text{Q}\mathopen{} \Bigg(\mathopen{}\dfrac{\Delta}{\sigma_{\mathcal{R}_{Y,\text{FA1}}}} \mathopen{}\Bigg)
    \label{eq:P_fa_si}
\end{align}
where $\sigma^2_{\mathcal{R}_{Y,\text{FA1}}}=N/2$. 
For the partial-correlation inflicted false alarms, we shall obtain an asymptotic distribution of $\mathcal{R}_\text{D}(k)$ by employing Slutsky's lemma \cite[Lemma 2.8]{asymptotic_statistics}. Using the model of the codewords from \eqref{eq:codeword_def}, we obtain 
\begin{IEEEeqnarray}{l}
    \mathcal{R}_\text{D}(k) = \Re\lefto[\sqrt{P}\mathbf{p}^\text{H} \mathbf{D}^{(N)} \righto] =\Re\lefto[ \dfrac{\sqrt{PN}\mathbf{p}^\text{H} \mathbf{\tilde{D}}^{(N)}}{\vectornormbig{\mathbf{\tilde{D}}^{(N)}}}    \righto].
\end{IEEEeqnarray}
Here, $\mathbf{p}^\text{H}$ is the transposed conjugate of the $N$-dimensional vector of preamble symbols $p_j$.
Next, we show that $\mathcal{R}_{\text{D}}(k)$ is well-approximated by Gaussian random variable for large $N$. To this end, let  $ Z = \Re\lefto[\sqrt{P} \mathbf{p}^\text{H} \mathbf{\tilde{D}}^{(N)}\righto]$. Then, it follows that $Z$ is a Gaussian random variable with zero mean and variance
\begin{IEEEeqnarray}{rCl}
    \mu_Z &=& \E \mathopen{}\left[\Re\lefto[\sqrt{P} \mathbf{p}^\text{H} \mathbf{\tilde{D}}^{(N)} \righto]\right] =0; \\
    \sigma_Z^2 & =& \Var \mathopen{}\left[ \Re \lefto[ \sqrt{P} \mathbf{p}^\text{H} \mathbf{\tilde{D}}^{(N)}\righto]\right] \mathopen{}=  \frac{1}{2}P N .\IEEEeqnarraynumspace
    \label{eq:var_z}
\end{IEEEeqnarray}
Since the law of large numbers implies that $\vectornormbig{\mathbf{\tilde{D}}^{(N)}}/\sqrt{N}\scriptedarrow{p}1$, it follows from Slutsky's lemma \cite[Lemma 2.8]{asymptotic_statistics} that 
\begin{align}
    \mathcal{R}_\text{D}(k)=\Re\lefto[\sqrt{P}\mathbf{p}^\text{H}\mathbf{D}^{(N)}\righto] \scriptedarrow{d} Z \sim \mathcal{N}(\mu_Z,\sigma_Z^2)
     \label{eq:Slutsky_proof}
\end{align}
where $\scriptedarrow{d} \mbox{and } \scriptedarrow{p}$ denote convergence in distribution and convergence in probability, respectively.
Therefore, as $N\rightarrow \infty$, $\mathcal{R}_{Y,\tau-k}$ converges to a Gaussian random variable with mean and variance 
\begin{IEEEeqnarray}{rCl}
    \mu_{\mathcal{R}_{Y,\text{FA2}}} (k) &=& \sqrt{\alpha P}\mathcal{R}_\text{p}^\text{SI}(k) ;\\
    \sigma_{\mathcal{R}_{Y,\text{FA2}}} ^2 (k) &=& \Var\mathopen{}\left[\mathcal{R}_W(k) + \sqrt{(1-\alpha)P} \mathcal{R}_\text{D}(k)\right]  
    = \sigma_{\mathcal{R}_{Y,\text{FA1}}}^2 + (1-\alpha) \Var\Bigg[\Re\bigg(\sqrt{P}\sum_{j=k}^{N-1} p_j^* D_{j-k}\bigg)\Bigg] 
    = \dfrac{N}{2} + \frac{1}{2}(1-\alpha)(N-k)P.\IEEEeqnarraynumspace\label{eq:sigmaPsi}
\end{IEEEeqnarray}
Here, \eqref{eq:sigmaPsi} follows from \eqref{eq:var_z}, \eqref{eq:Slutsky_proof}, \eqref{eq:SI_sigma_h1}, and because the covariance matrix of $\mathbf{D}^{(N)}$ is the identity matrix.\footnote{Spherical symmetry implies that the vector $\mathbf{D}^{(N)}$ in \eqref{eq:codeword_def} has zero mean with entries being uncorrelated and with entries having unit variance. The latter follows because  $\sum_{i=1}^N\E[(D_i)^2] = N \E\mathopen{}\Big[\frac{\sum_{i=1}^N (\tilde D_i)^2}{||\mathbf{\tilde D}^{(N)} ||^2 }\Big] = N$.}
Therefore, a partial-correlation inflicted false alarm at time $\tau-k$ is 
\begin{align}
    \Pr\left[\mathcal{R}_{Y,\tau-k}>\Delta\right] &\approx \text{Q}\Bigg( \dfrac{\Delta-\mu_{\mathcal{R}_{Y,\text{FA2}}} (k)} {\sigma_{\mathcal{R}_{Y,\text{FA2}}}(k)}\Bigg).
\end{align}
Applying the union bound as in \eqref{eq:P_fa2}, the probability of having a partial-correlation inflicted false alarm for the superimposed case is upper-bounded as
\begin{align}
    \Pr\mathopen{}\left[{\mathcal{E}_\text{FA2}} \right] \leq \sum_{k=1}^{N-1} \text{Q}\Bigg( \dfrac{\Delta-\mu_{\mathcal{R}_{Y,\text{FA2}}} (k)} {\sigma_{\mathcal{R}_{Y,\text{FA2}}}(k)}\Bigg).
    \label{eq:p_fa2_si}
\end{align}
%
Finally, for the misdetection event, the correlation metric can also be approximated as a Gaussian variable. Therefore, the probability of a misdetection is approximated by
\begin{align}
    \Pr \mathopen{} \left[ \mathcal{E}_\text{MD}\right] &\approx \text{Q} \mathopen{}\Bigg(\dfrac{\mu_{\mathcal{R}_{Y,\tau}} - \Delta}{\sigma_{\mathcal{R}_{Y,\tau}}} \Bigg). \label{eq:p_md_si}
\end{align}
Here, the mean and variance are
\begin{IEEEeqnarray}{cl}
    \mu_{\mathcal{R}_{Y,\tau}}&=\sqrt{\alpha P} N + \sqrt{1-\alpha} \mu_Z = \sqrt{\alpha P} N \label{eq:SI_mu_h1}\\
    \sigma_{\mathcal{R}_{Y,\tau}}^2 &= \sigma^2_{\mathcal{R}_{Y,\text{FA1}}} + (1-\alpha) \Var\left[ \Re \lefto[ \sqrt{P} \mathbf{p}^\mathrm{H} \mathbf{D}^{(N)} \righto]\right]  \label{eq:SI_sigma_h1}\\
    &= \frac{N}{2} + (1-\alpha) \sigma^2_Z  = \frac{N}{2} + \frac{1}{2}(1-\alpha) P N  \IEEEeqnarraynumspace\label{eq:var_Rytau}
\end{IEEEeqnarray} 
where \eqref{eq:var_Rytau} follows because the covariance matrix of $\mathbf{D}^{(N)}$ is the identity matrix.
\begin{figure} [ht]
    \centering
    \includegraphics [width = 0.75\columnwidth] {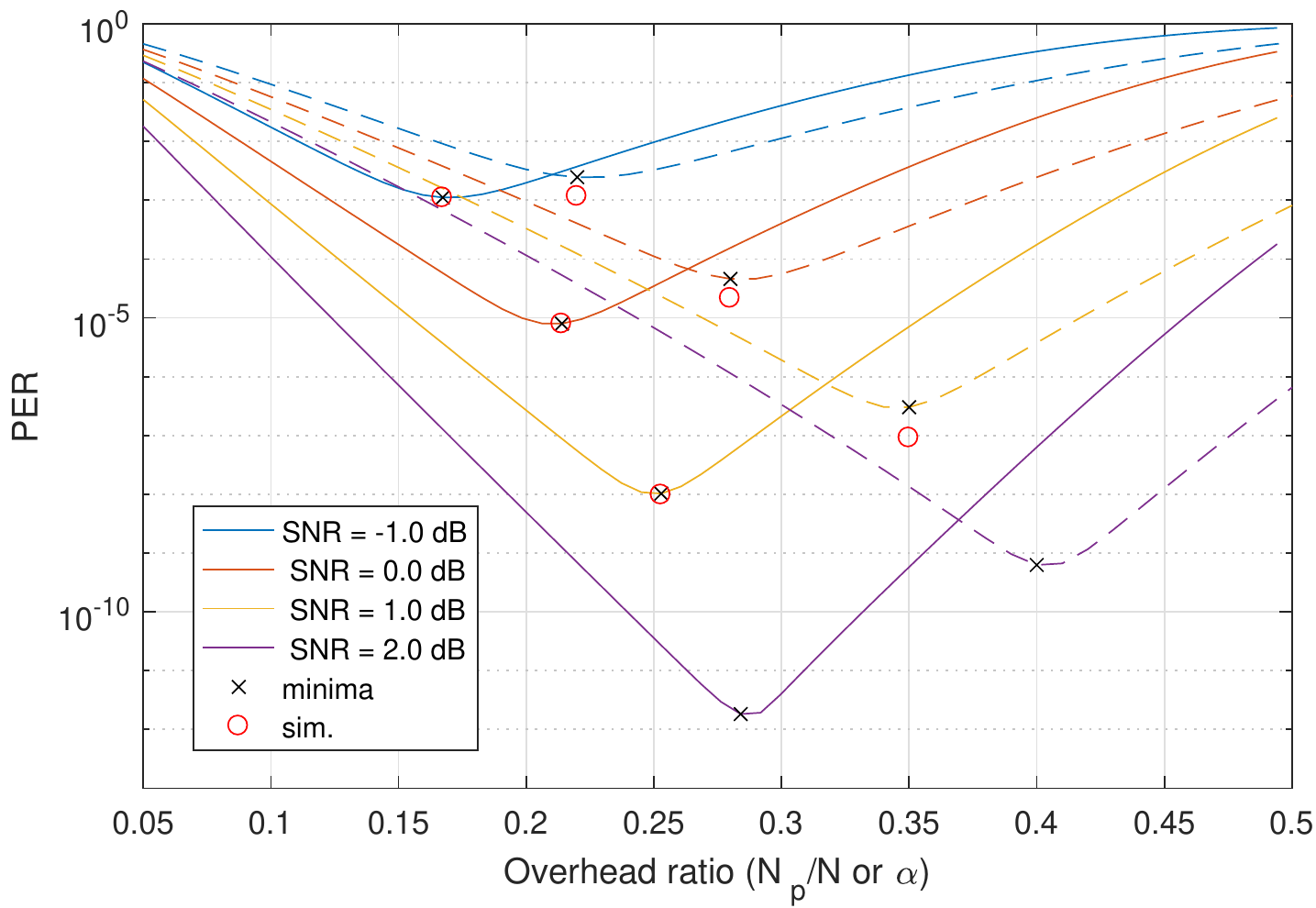}
    \caption{Upper bounds/approximations of PER for the preamble case (solid lines) and superimposed case (dashed lines) for different SNR values. Black dots and red circles represent minima of the upper bounds/approximations and simulated PER values, respectively. A packet carrying $b=128$ bits transmitted over $N=257$ channel uses is considered and a recovery time of $t_r=283$ channel uses is assumed ($10\%$ higher than $N$). For $\text{PER}<10^{-8}$, computational complexity precludes accurate results.} 
    \label{fig:optimiz_pre_si_alpha_Pe} 
\end{figure}
\begin{figure} [ht]
    \centering
    \includegraphics [width = 0.75 \columnwidth] {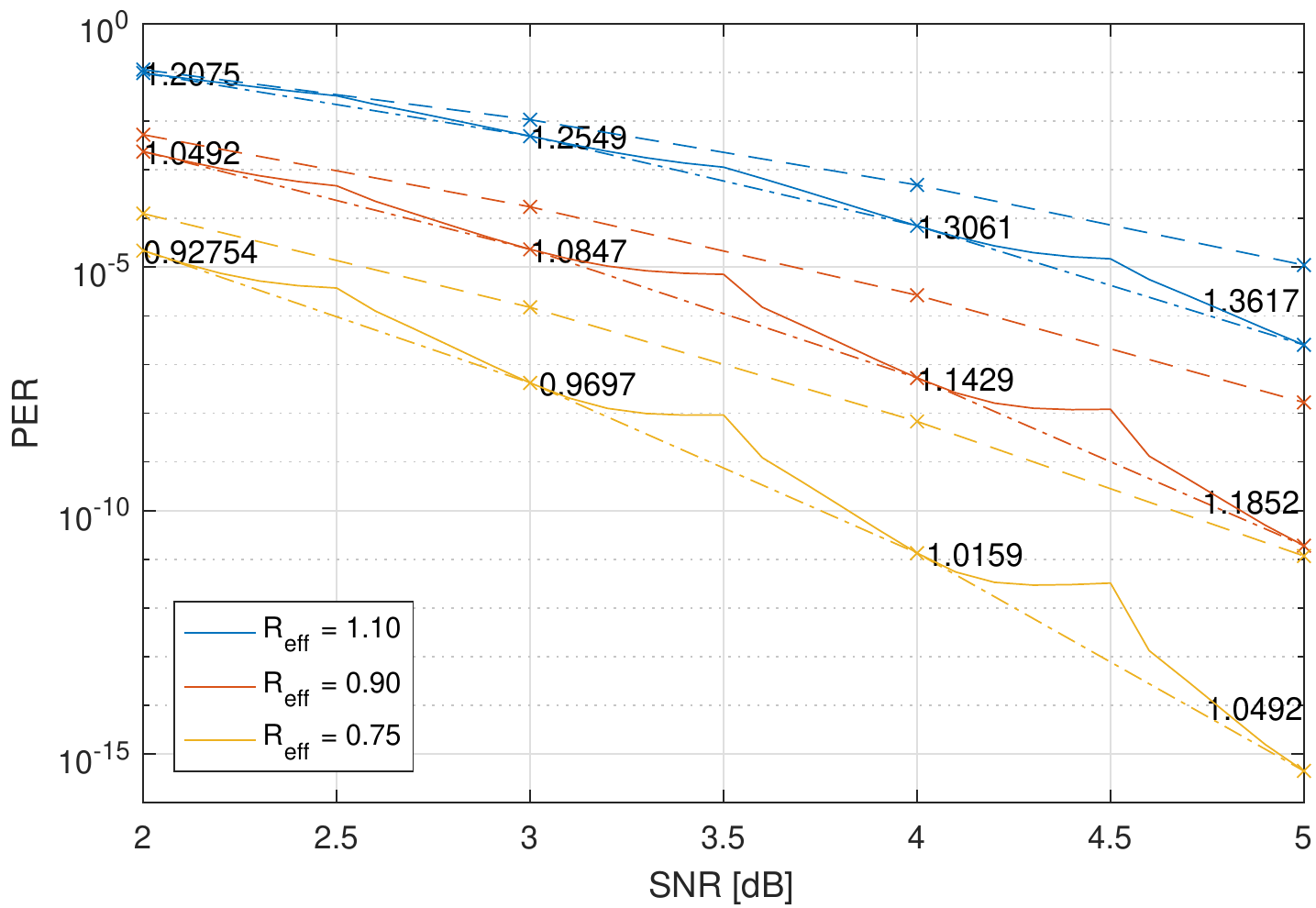}
    \caption{PER for three effective rates ($b/N$, $b=128$, $N$ varies) for optimal preamble size (dotted and dashed lines), optimal superimposed sequence (dashed lines) and pragmatic approach of adaptive coding rate for the preamble case (solid line), where the coding rate for the data is changed every \SI{1}{dB} interval centered in the optimal computed rate for each integer SNR point, and labeled by the numbers next to it. } 
    \label{fig:Pe_v_SNR_prag}
\end{figure}
\section{Numerical results and Conclusion}

In this section, we plot our analytical bounds and approximations and compare them to simulations of the PER. In the simulations of the PER, we use the approximation \eqref{eq:error_finite_blocklength} to compute the decoding error probability.  

For the time-multiplexed preamble, we let $ \overline{\mathcal{P}_{e}^\text{pre}}(\Delta,N_\text{p},N,P)$ denote the upper bound on the PER obtained by summing \eqref{eq:P_fa2} and \eqref{eq:misdetect}, and $\epsilon_\text{D}(N-N_\text{p}, P)$. We are interested in solving the optimization problem
\begin{IEEEeqnarray}{rCll}
& \underset{ \substack{ N_\text{p}\in\{1,\ldots,N-1\}\\ \Delta\geq 0} }{\text{minimize}}
&\quad & \overline{\mathcal{P}_{e}^\text{pre}}(\Delta,N_\text{p},N,P).
\label{eq:min_pe_pre}
\end{IEEEeqnarray}
Note that the decoding error probability depends on $N_\text{p}$ and $P$ and the detection error probability depends on $\Delta, N_\text{p}$, and $P$. 

For the superimposed sequence, we let $\overline{\mathcal{P}_{e}^\text{SI}}(\Delta,\alpha,N,P)$ be the approximation of the PER obtained by summing \eqref{eq:P_fa_si}, \eqref{eq:p_fa2_si}, \eqref{eq:p_md_si}, and $\epsilon_\text{D}(N, (1-\alpha)P)$. Our objective is to solve the optimization problem 
\begin{IEEEeqnarray}{rCll}
& \underset{\substack{\alpha\in(0,1)\\ \Delta\geq 0}}{\text{minimize}}
&\quad & \overline{\mathcal{P}_{e}^\text{SI}}(\Delta,\alpha,N,P).
\label{eq:minimiz_si}
\end{IEEEeqnarray}

The tradeoff between detection and decoding  is shown in Fig.~\ref{fig:optimiz_pre_si_alpha_Pe}, which also depicts the solutions to the optimization problems above.  For all points of the solid and dashed curves, $\Delta$ is optimized. We observe that the optimal overhead ratios depend on the SNR and the target reliability, and that the superimposed structure achieves its minimum error probability at a higher overhead ratio. Furthermore, the PER achieved with superimposed detection sequences also experiences a degradation in terms of minimum PER because a larger fraction of resources are spent on detection overhead. 

The superimposed structure, however, offers enhanced adaptivity. For the system model considered, where the receiver and transmitter are aware of the SNR, the preamble and superimposed structures follow the optimal regimes with the variation of $N_\text{p}$ and $\alpha$, respectively. For the preamble case, this requires that the codewords are encoded with different rates. A pragmatic approach of what can be achieved is shown in Fig.~\ref{fig:Pe_v_SNR_prag} in solid lines, where a distinct codebook is required for each \SI{1}{dB} SNR interval. For comparison, the dashed lines and the dashed-dotted lines indicate the superimposed and the ideal preamble optimal regimes, respectively.


\bibliographystyle{IEEEtran}
\bibliography{mybib}

\begin{thebibliography}{10}
\providecommand{\url}[1]{#1}
\csname url@samestyle\endcsname
\providecommand{\newblock}{\relax}
\providecommand{\bibinfo}[2]{#2}
\providecommand{\BIBentrySTDinterwordspacing}{\spaceskip=0pt\relax}
\providecommand{\BIBentryALTinterwordstretchfactor}{4}
\providecommand{\BIBentryALTinterwordspacing}{\spaceskip=\fontdimen2\font plus
\BIBentryALTinterwordstretchfactor\fontdimen3\font minus
  \fontdimen4\font\relax}
\providecommand{\BIBforeignlanguage}[2]{{%
\expandafter\ifx\csname l@#1\endcsname\relax
\typeout{** WARNING: IEEEtran.bst: No hyphenation pattern has been}%
\typeout{** loaded for the language `#1'. Using the pattern for}%
\typeout{** the default language instead.}%
\else
\language=\csname l@#1\endcsname
\fi
#2}}
\providecommand{\BIBdecl}{\relax}
\BIBdecl

\bibitem{pkt_structure_urllc}
B.~Lee, S.~Park, D.~J. Love, H.~Ji, and B.~Shim, ``Packet structure and
  receiver design for low latency wireless communications with ultra-short
  packets,'' \emph{IEEE Trans. Commun.}, vol.~66, no.~2, Feb. 2018.

\bibitem{towards_massiveURLLC_short_pkts}
G.~Durisi, T.~Koch, and P.~Popovski, ``Towards massive, ultrareliable, and
  low-latency wireless communication with short packets,'' \emph{Proc. IEEE},
  vol. 104, no.~9, Sep. 2016.

\bibitem{URLLC_principles_magazine}
P.~Popovski, J.~J. Nielsen, {\v C}.~Stefanovi{\' c}, E.~de~Carvalho,
  E.~Str{\"o}m, K.~F. Trillingsgaard, A.-S. Bana, D.~M. Kim, R.~Kotaba,
  J.~Park, and R.~B. Sørensen, ``Wireless access for ultra-reliable
  low-latency communication {(URLLC)}: Principles and building blocks,'' 2017,
  \texttt{arXiv:1708.07862 [cs.IT]}.

\bibitem{optimum_frame_sync_massey}
J.~L. Massey, ``Optimum frame synchronization,'' \emph{IEEE Trans. Commun.},
  vol. COM-20, no.~2, Apr. 1972.

\bibitem{comm_strong_async_tcham}
A.~Tchamkerten, V.~Chandar, and G.~W. Wornell, ``Communication under strong
  asynchronism,'' \emph{IEEE Trans. Inf. Theory}, vol.~55, no.~10, Oct. 2009.

\bibitem{asynchr_polyanskiy}
Y.~Polyanskiy, ``Asynchronous communications: Exact synchronization,
  universality, and dispersion,'' \emph{IEEE Trans. Inf. Theory}, vol.~59,
  no.~3, Mar. 2013.

\bibitem{codeword_or_noise}
N.~Weinberger and N.~Merhav, ``Codeword or noise? {Exact} random coding
  exponents for joint detection and decoding,'' \emph{IEEE Trans. Inf. Theory},
  vol.~60, no.~9, Sep. 2014.

\bibitem{differential_preamble_detection}
S.~Nagaraj, S.~Khan, C.~Schlegel, and M.~V. Burnashev, ``Differential preamble
  detection in packet-based wireless networks,'' \emph{IEEE Trans. Wireless
  Commun.}, vol.~8, no.~2, Feb. 2009.

\bibitem{superimposed_ofdm_wlan}
Y.~Wang, J.~Oostveen, A.~Filippi, and S.~Wesemann, ``A novel preamble scheme
  for packet-based {OFDM} {WLAN},'' in \emph{Proc. IEEE Wireless Commun. and
  Networking Conf.}, Kowloon, China, Mar. 2007.

\bibitem{low_latency_UR5g_finite_blocklength}
J.~{\"O}stman, G.~Durisi, E.~G. Str{\"o}m, J.~Li, H.~Sahlin, and G.~Liva,
  ``Low-latency ultra-reliable {5G} communications: Finite-blocklength bounds
  and coding schemes,'' in \emph{Proc. IEEE Conf. on Sys., Commun. and Coding},
  Hamburg, Germany, Feb. 2017.

\bibitem{finite_blocklength_polyanskiy}
Y.~Polyanskiy, H.~V. Poor, and S.~Verdú, ``Channel coding rate in the finite
  blocklength regime,'' \emph{IEEE Trans. Inf. Theory}, vol.~56, no.~5, Apr.
  2010.

\bibitem{liva_mismatched_csi}
G.~Liva, G.~Durisi, M.~Chiani, S.~S. Ullah, and S.~C. Liew, ``Short codes with
  mismatched channel state information: A case study,'' 2017,
  \texttt{arXiv:1705.05528 [cs.IT]}.

\bibitem{chu_polyphase_codes}
D.~C. Chu, ``Polyphase codes with good periodic correlation properties,''
  \emph{IEEE Trans. Inf. Theory}, vol.~18, no.~4, Jul. 1972.

\bibitem{LTE_for_4G_farooq_khan}
F.~Khan, \emph{LTE for 4G Mobile Broadband: Air Interface Technologies and
  Performance}.\hskip 1em plus 0.5em minus 0.4em\relax {ISBN}: 9780521882217:
  Cambridge University Press, 2009.

\bibitem{MIT_intro}
\BIBentryALTinterwordspacing
A.~V. Oppenheim and G.~C. Verghese, ``6.011 {Introduction} to communication,
  control and signal processing,'' Massachusetts Institute of Technology: MIT
  OpenCourseWare. License: Creative Commons BY-NC-SA, Spring 2010. [Online].
  Available: \url{https://ocw.mit.edu}
\BIBentrySTDinterwordspacing

\bibitem{asymptotic_statistics}
A.~van~der Vaart, \emph{Asymptotic Statistics}.\hskip 1em plus 0.5em minus
  0.4em\relax {ISBN}: 978-0-521-49603-2: Cambridge University Press, 2007.

\end{thebibliography}
\end{document}